\documentstyle[seceq,epsf,preprint]{ptptex}

\newcommand{\bra}[1]{\langle {#1} |}     
\newcommand{\ket}[1]{| {#1} \rangle}     
\newcommand{\kket}[1]{| {#1} \rangle\!\rangle}     
\newcommand{\rket}[1]{| {#1} )}     
\newcommand{\wtilde}[1]{\widetilde{#1}} 

\title{
A Possible Boson Realization of the $so(4)$- and\\
the $so(3,1)$-Algebra
}
\subtitle{In Relation to the Runge-Lenz-Pauli Vector
}    

\author{
Seiya {\sc Nishiyama}$^{1}$, 
Constan\c{c}a {\sc Provid\^encia},$^{2}$\\
Jo\~ao da {\sc Provid\^encia}$^{2}$, Yasuhiko {\sc Tsue}$^{1}$ 
and Masatoshi {\sc Yamamura}$^{3}$
}

\inst{
$^{1}$Physics Division, Faculty of Science, Kochi University, Kochi 780-8520, 
Japan\\
$^{2}$Departamento de Fisica, Universidade de Coimbra, P-3000 Coimbra, 
Portugal\\
$^{3}$Faculty of Engineering, Kansai University, Suita 564-8680, Japan
}

\recdate{
\today
}

\abst{
As natural extensions of the boson realizations of the $su(2)$- and the 
$su(1,1)$-algebra, the $so(4)$- and the $so(3,1)$-algebras are presented 
in the form of boson realizations with four kinds of boson operators. 
For each algebra, two forms are discussed. 
One is constructed in terms of two sets of the boson operators 
which play a role of spherical tensor with rank $1/2$. 
The other is based on the ranks $1$ and $0$. 
As a possible application, the Runge-Lenz-Pauli vector, which is 
famous in the hydrogen atom, is derived with some aspects. 
}

\begin{document}

\maketitle

\section{Introduction}
It is hardly necessary to mention, but boson realization of Lie algebras has 
occupied historically a part of central positions in theoretical studies of 
nuclear dynamics. For example, we can find its fact concretely in a 
classical review article by Klein and Marshalek.\cite{1}
One of the recent interests of Lie algebraic approach to nuclear dynamics 
is related to the $so(4)$-algebra. With the aid of this algebra, 
nuclear dynamics induced by the pairing plus quadrapole interaction can be 
described schematically.\cite{2}
Of course, the $so(4)$-algebraic approach to nuclear dynamics 
has been already investigated by several authors.\cite{3} 
If we stand on the Lie algebraic viewpoint, we cannot forget a famous 
historical story. 
The $so(4)$-algebra gives us an interesting viewpoint for the understanding 
of hydrogen atom. 
Concerning this viewpoint, the Runge-Lenz-Pauli vector\cite{4} is fundamental. 
With the aid of this algebra, we are able to obtain a quite transparent 
understanding of the hydrogen atom. In associating with the $so(4)$-algebra, 
scattering of electron in the Coulomb field induced by proton 
(scattering problem) may be described in terms of the $so(3,1)$-algebra. 
The above is a historical interest of the $so(4)$-algebra. 
In order to respond to the above-mentioned situation, it may be significant 
to investigate the boson realization of the $so(4)$-algebra and its associated 
$so(3,1)$-algebra. 
We can describe nuclear dynamics related to the $so(4)$-algebra in terms of 
various techniques such as the use of the boson coherent and the boson 
squeezed states. 
These have played a central role for the description of 
many-boson systems. 

It is well known that the orthogonal set characterizing the $su(2)$-algebra 
is essentially specified by two quantum numbers. 
Therefore, a possible boson realization of the $su(2)$-algebra can be 
formulated in terms of two kinds of boson operators. 
This is identical with the Schwinger boson representation.\cite{5}
The case of the $su(1,1)$-algebra is in the same situation. 
It is characteristic that the two kinds of boson creation (annihilation) 
operators form a spherical tensor with rank $1/2$ (spinor). 
We borrow this idea. The orthogonal set for the $so(4)$-algebra is specified 
by four quantum numbers, and then, the boson realization of the 
$so(4)$-algebra may be obtained in terms of four kinds of boson operators. 
These enable us to construct the boson realization in two forms. 
First consists of two sets. Each set is composed of two kinds of boson 
which play a role of spherical tensor with rank $1/2$. 
Second is constructed in terms of three kinds of bosons plus one 
kind of boson. 
Three kinds of boson creation (annihilation) operators form a spherical 
tensor with rank $1$ (vector) and one kind of boson creation (annihilation) 
operators forms a spherical tensor with rank $0$ (scalar). 
The above is our starting idea for constructing the $so(4)$-algebra. 
The case of the $so(3,1)$-algebra is also in the same situation.

Following the above-mentioned idea, we present the boson realization in which 
six operators forming the $so(4)$-algebra are expressed in terms of 
bilinear form for the four kinds of boson operators. 
We can also give the case of the $so(3,1)$-algebra. Each algebra has two 
forms of the boson realization. 
In these forms, the $so(4)$-algebra in two sets of the spinors is the 
most interesting in our present knowledge. 
This case can be reduced to the hydrogen atom and the pairing plus 
quadrapole interaction. 
We can show that this boson realization is equivalent to the form 
developed by the present authors,\cite{6} which is referred to as (A). 
The form given in (A) is a simple application of the general form 
for the $su(M+1)$- and the $su(N,1)$-algebra presented by the present 
authors.\cite{7} 
Technical points appearing in \S\S 4$\sim$ 6 are discussed in Ref.\citen{8}, 
which is referred to as (B). 
In subsequent papers, we will report the results on the pairing plus 
quadrapole interaction. 
In the present paper, we discuss the problem of the hydrogen atom 
and the scattering problem. 
The basic idea is, in some sense, to transcribe the form presented 
in terms of four kinds of bosons into the form expressed by certain 
parameters. 
With the aid of this transcription, we obtain certain orthogonal set in four 
dimension space which can be expressed in terms of the Laguerre polynomial 
and the $D$-function. 
And, further, under a certain condition, the hydrogen atom is described. 
Of course, the Runge-Lenz-Pauli vector is derived. 
In the case of the scattering problem based on the $so(3,1)$-algebra, 
we also obtain the Runge-Lenz-Pauli vector. 

In the next section, a general framework of the $so(4)$- and the 
$so(3,1)$-algebra is recapitulated as a preliminary argument. 
In \S 3, the boson realization for the case of the 
$su(2)\otimes su(2)$-algebra is presented. 
Section 4 is devoted to giving the form based on the vector and the scalar 
bosons. 
In \S\S 5 and 6, the $so(3,1)$-algebra in two forms are presented. 
Finally, in \S 7, the approach to the description of the hydrogen atom 
and the scattering problem 
is given.

\section{Preliminary argument}

First, we recapitulate the basic part of the $so(4)$-algebra. 
In this paper, six operators composing this algebra are denoted as 
${\hat L}_{\pm,0}$ and ${\hat M}_{\pm,0}$. The hermitian property 
is listed up as 
\begin{equation}\label{2-1}
{\hat L}_-^*={\hat L}_+ \ , \qquad
{\hat L}_0^*={\hat L}_0 \ , \qquad
{\hat M}_-^*={\hat M}_+ \ , \qquad
{\hat M}_0^*={\hat M}_0 \ .
\end{equation}
The commutation relations are given as 
\begin{subequations}\label{2-2}
\begin{eqnarray}
& &[\ {\hat L}_+\ , \ {\hat L}_-\ ]=2{\hat L}_0 \ , \qquad
[\ {\hat L}_0\ , \ {\hat L}_\pm \ ]=\pm{\hat L}_\pm \ , 
\label{2-2a}\\
& &[\ {\hat L}_\pm \ , \ {\hat M}_\pm\ ]=0 \ , \qquad
[\ {\hat L}_\pm\ , \ {\hat M}_0 \ ]=\mp{\hat M}_\pm \ , \qquad
[\ {\hat L}_\pm\ , \ {\hat M}_\mp \ ]=\pm 2{\hat M}_0 \ , \nonumber\\
& &[\ {\hat L}_0 \ , \ {\hat M}_\pm \ ]=\pm{\hat M}_\pm \ , \qquad
[\ {\hat L}_0\ , \ {\hat M}_0 \ ]=0 \ , 
\label{2-2b}\\
& &[\ {\hat M}_+\ , \ {\hat M}_-\ ]=2{\hat L}_0 \ , \qquad
[\ {\hat M}_0\ , \ {\hat M}_\pm \ ]=\pm{\hat L}_\pm \ . 
\label{2-2c}
\end{eqnarray}
\end{subequations}
The set $({\hat L}_{\pm,0})$ forms the $su(2)$-algebra.

The orthogonal set for the $so(4)$-algebra is specified by four 
quantum numbers, and then, we must prepare four hermitian operators 
which are mutually commuted. 
As for them, for example, we can choose ${\hat {\mib \Gamma}}_1$, 
${\hat {\mib \Gamma}}_2$, ${\hat L}_0$ and ${\hat M}_0$, where 
${\hat {\mib \Gamma}}_1$ and ${\hat {\mib \Gamma}}_2$ are defined as 
\begin{subequations}
\begin{equation}\label{2-3}
{\hat {\mib \Gamma}}_1=(1/4)({\hat {\mib L}}^2+{\hat {\mib M}}^2) \ , \qquad
{\hat {\mib \Gamma}}_2=(1/2){\hat {\mib L}}\cdot{\hat {\mib M}}\ 
(=(1/2){\hat {\mib M}}\cdot{\hat {\mib L}}) \ . 
\end{equation}
It should be noted that ${\hat {\mib \Gamma}}_1$ and ${\hat {\mib \Gamma}}_2$ 
play a role of the Casimir operators for the $so(4)$-algebra : 
\begin{equation}\label{2-3a}
[\ {\hat {\mib \Gamma}}_1 \ {\rm and}\ {\hat {\mib \Gamma}}_2\ , \ 
{\rm any\ of\ }({\hat L}_{\pm,0}\ , \ {\hat M}_{\pm,0})\ ]=0 \ . 
\end{equation}
\end{subequations}
In this paper, we use the notation 
${\mib A}\cdot{\mib B}$ for ${\hat A}_{\pm,0}$ and ${\hat B}_{\pm,0}$ as
\begin{equation}\label{2-4}
{\mib A}\cdot{\mib B}=A_0B_0+(1/2)(A_-B_+ + A_+B_-) \ . 
\end{equation}
If ${\mib A}={\mib B}$, ${\mib A}\cdot{\mib B}$ is denoted as ${\mib A}^2$. 
Of course, the above framework is based on the condition 
\begin{equation}\label{2-5}
{\hat {\mib \Gamma}}_2=(1/2){\hat {\mib L}}\cdot{\hat {\mib M}} \neq 0 \ .
\end{equation}
Later, we will investigate the framework obeying the condition 
${\hat {\mib \Gamma}}_2=(1/2){\hat {\mib L}}\cdot{\hat {\mib M}} =0$. 

For the framework obeying the condition (\ref{2-5}), it may be 
convenient to describe the $so(4)$-algebra in terms of 
$({\hat \Lambda}_{\pm,0}\ , \ {\hat K}_{\pm,0})$ defined as 
\begin{subequations}\label{2-6}
\begin{eqnarray}
& &{\hat \Lambda}_{\pm,0}=(1/2)({\hat L}_{\pm,0}+{\hat M}_{\pm,0}) \ , 
\label{2-6a}\\
& &{\hat K}_{\pm,0}=(1/2)({\hat L}_{\pm,0}-{\hat M}_{\pm,0}) \ .  
\label{2-6b}
\end{eqnarray}
\end{subequations}
The commutation relation (\ref{2-2}) leads us to 
\begin{subequations}\label{2-7}
\begin{eqnarray}
& &[\ {\hat \Lambda}_+\ , \ {\hat \Lambda}_-\ ]=2{\hat \Lambda}_0 \ , \qquad
[\ {\hat \Lambda}_0\ , \ {\hat \Lambda}_\pm\ ]=\pm{\hat \Lambda}_\pm \ , 
\label{2-7a}\\
& &[\ {\hat K}_+\ , \ {\hat K}_-\ ]=2{\hat K}_0 \ , \qquad
[\ {\hat K}_0\ , \ {\hat K}_\pm\ ]=\pm{\hat K}_\pm \ , 
\label{2-7b}\\
& &[\ {\rm any\ of\ }{\hat \Lambda}_{\pm,0}\ , 
\ {\rm any\ of\ }{\hat K}_{\pm,0} ]=0 \ . 
\label{2-7c}
\end{eqnarray}
\end{subequations}
Clearly, each obeys the $su(2)$-algebra, which implies that the 
$so(4)$-algebra is reduced to the $su(2)\otimes su(2)$-algebra. 
For the relation (\ref{2-3}), we have 
\begin{equation}\label{2-8}
{\hat {\mib \Gamma}}_1+{\hat {\mib \Gamma}}_2={\hat {\mib \Lambda}}^2 \ , 
\qquad 
{\hat {\mib \Gamma}}_1-{\hat {\mib \Gamma}}_2={\hat {\mib K}}^2 \ . 
\end{equation}
Therefore, ${\hat {\mib \Lambda}}^2$ and ${\hat {\mib K}}^2$ play a role of 
the Casimir operators for the $su(2)\otimes su(2)$-algebra and the 
mutually commutable four operators are $({\hat {\mib \Lambda}}^2 , 
{\hat \Lambda}_0, {\hat {\mib K}}^2 , {\hat K}_0)$\break
or 
$({\hat {\mib \Lambda}}^2 , {\hat {\mib K}}^2 , {\hat {\mib L}}^2 , 
{\hat L}_0)$. 
A concrete example will be presented in \S 3.

Next, we investigate the case governed by the condition 
\begin{equation}\label{2-9}
{\hat {\mib \Gamma}}_2=(1/2){\hat {\mib L}}\cdot{\hat {\mib M}}=0 \ . 
\end{equation}
In this case, the relations (\ref{2-6}) and (\ref{2-7}) give us 
\begin{equation}\label{2-10}
{\hat {\mib \Gamma}}_1={\hat {\mib \Lambda}}^2={\hat {\mib K}}^2 \ . 
\end{equation}
Then, the mutually commutable operators are reduced to 
$({\hat {\mib \Gamma}}_1,{\hat \Lambda}_0, {\hat K}_0)$ or  
$({\hat {\mib \Gamma}}_1,{\hat {\mib L}}^2, {\hat L}_0)$. 
This implies that we must find one more hermitian operator which commutes 
with the above. 
In general framework, it may be impossible to present the explicit 
form as a function of $({\hat \Lambda}_{\pm,0} , {\hat K}_{\pm,0})$. 
We can show its example in a concrete case. 
In \S 4, the example will be discussed.

Our next concern is to investigate the $so(3,1)$-algebra, in which the set 
$({\hat M}_{\pm,0})$ in the $so(4)$-algebra is 
replaced with the set $({\wtilde M}_{\pm,0})$ obeying 
\begin{equation}\label{2-11}
[\ {\wtilde M}_+\ , \ {\wtilde M}_-\ ]=-2{\hat L}_0 \ , \qquad
[\ {\wtilde M}_0\ , \ {\wtilde M}_\pm\ ]=\mp {\hat L}_\pm \ . 
\end{equation}
The commutation relations (\ref{2-2a}) and (\ref{2-2b}) conserve their forms. 
It should be noted that the sign on the right-hand side of the relation 
(\ref{2-11}) is inverse from that of the relation (\ref{2-2c}). 
In a form analogous to the form (\ref{2-3}), we introduce two 
hermitian operators ${\wtilde {\mib \Gamma}}_1$ and 
${\wtilde {\mib \Gamma}}_2$, which play a role of the Casimir operators in 
the $so(3,1)$-algebra : 
\begin{subequations}
\begin{eqnarray}
& &{\wtilde {\mib \Gamma}}_1=(1/4)({\hat {\mib L}}^2-{\wtilde {\mib M}}^2) \ , 
\qquad 
{\wtilde {\mib \Gamma}}_2=(1/2){\hat {\mib L}}\cdot{\wtilde {\mib M}} \ 
(=(1/2){\wtilde {\mib M}}\cdot{\hat {\mib L}})\ , 
\label{2-12}\\
& &[\ {\wtilde {\mib \Gamma}}_1 \ {\rm and}\ {\wtilde {\mib \Gamma}}_2\ , \ 
{\rm any\ of\ }({\hat L}_{\pm,0}\ , \ {\wtilde M}_{\pm,0})\ ]=0 \ . 
\label{2-12a}
\end{eqnarray}
\end{subequations}
In this case, also, there exist two cases ${\wtilde {\mib \Gamma}}_2 
\neq 0$ and ${\wtilde {\mib \Gamma}}_2=0$. 
Until the present stage, the formalism is completely analogous to the case 
of the $so(4)$-algebra. 
However, in the present general framework, further discussion 
may be impossible. 
In concrete examples, it may be possible and in \S\S 5 and 6, we will present 
the examples.

\section{An example obeying the condition ${\hat {\mib \Gamma}}_2\neq 0$ 
in the $so(4)$-algebra}

The example discussed in this section is essentially the same as that 
presented in (A). However, in (A), the terminology 
``$so(4)$-algebra" has not been used and the notations adopted in (A) 
are different from the present ones. 
Since the orthogonal set for the $so(4)$-algebra is specified in 
terms of four quantum numbers, a concrete description may be 
carried out in the framework of four kinds of boson operators. 
We denote them as $({\hat c}_\pm^* , {\hat c}_\pm)$ and 
$({\hat d}_\pm^* , {\hat d}_\pm)$. 
With the use of these boson operators, ${\hat \Lambda}_{\pm,0}$ and 
${\hat K}_{\pm,0}$ can be expressed as follows : 
\begin{subequations}\label{3-1}
\begin{eqnarray}
& &{\hat \Lambda}_+={\hat c}_+^*{\hat d}_+ \ , \qquad 
{\hat \Lambda}_-={\hat d}_+^*{\hat c}_+ \ , \qquad 
{\hat \Lambda}_0=(1/2)({\hat c}_+^*{\hat c}_+ -{\hat d}_+^*{\hat d}_+) \ , 
\label{3-1a}\\
& &{\hat K}_+={\hat c}_-^*{\hat d}_- \ , \qquad 
{\hat K}_-={\hat d}_-^*{\hat c}_- \ , \qquad 
{\hat K}_0=(1/2)({\hat c}_-^*{\hat c}_- -{\hat d}_-^*{\hat d}_-) \ . 
\label{3-1b}
\end{eqnarray}
\end{subequations}
The operators $({\hat c}_{\pm} , {\hat d}_{\pm})$ and 
$({\hat \Lambda}_{\pm,0}, {\hat K}_{\pm,0})$ correspond to 
$({\hat a}_\pm,{\hat b}_\pm)$ and $({\hat I}_{\pm,0},{\hat J}_{\pm,0})$ 
in (A), respectively. 
The above is identical with the Schwinger boson representation for the 
$su(2)$-algebra. 
The Casimir operator ${\hat {\mib \Gamma}}_2=(1/2){\hat {\mib L}}\cdot
{\hat {\mib M}}$ can be rewritten as 
${\hat {\mib \Gamma}}_2=(1/2)({\hat {\mib \Lambda}}^2-{\hat {\mib K}}^2)$, 
which does not vanish identically. 
The eigenstate for 
$({\hat {\mib \Lambda}}^2, {\hat \Lambda}_0 , {\hat {\mib K}}^2, {\hat K}_0)$ 
is given in the form 
\begin{eqnarray}\label{3-2}
& &\ket{\lambda\lambda_0;\kappa\kappa_0}
=({\hat \Lambda}_+)^{\lambda+\lambda_0}
({\hat K}_+)^{\kappa+\kappa_0}({\hat d}_+^*)^{2\lambda}({\hat d}_-^*)^{2\kappa}
\ket{0} \ , \nonumber\\
& &\ \ \lambda, \ \kappa=0,\ 1/2,\ 1,\ 3/2, \cdots \ , \nonumber\\
& &\ \ \lambda_0=-\lambda, \ -\lambda+1,\ \cdots ,\ \lambda-1,\ \lambda\ , 
\qquad
\kappa_0=-\kappa,\ -\kappa+1,\ \cdots ,\ \kappa-1,\ \kappa \ . \qquad
\end{eqnarray}
The eigenvalues of ${\hat {\mib \Lambda}}^2$, ${\hat \Lambda}_0$, 
${\hat {\mib K}}^2$ and ${\hat K}_0$ are given as 
$\lambda(\lambda+1)$, $\lambda_0$, $\kappa(\kappa+1)$ and $\kappa_0$, 
respectively. 
A normalization constant of the state (\ref{3-2}) is omitted, and hereafter, 
we omit the normalization constant for any state.

With the use of the Clebsch-Gordan coefficient, 
the eigenstate for $({\hat {\mib \Lambda}}^2,{\hat {\mib K}}^2, 
{\hat {\mib L}}^2, {\hat L}_0)$ can be expressed in the form 
\begin{subequations}\label{3-3}
\begin{eqnarray}
& &\ket{\lambda \kappa;ll_0}=\sum_{\lambda_0 \kappa_0}
\bra{\lambda\lambda_0 \kappa \kappa_0}ll_0\rangle 
\ket{\lambda\lambda_0;\kappa\kappa_0} \ , 
\nonumber\\
& &\ \ l=|\lambda-\kappa|,\ |\lambda-\kappa|+1,\ \cdots ,\ 
\lambda+\kappa-1,\ \lambda+\kappa \ , 
\nonumber\\
& &\ \ l_0=-l,\ -l+1,\ \cdots ,\ l-1,\ l \ . 
\label{3-3a}
\end{eqnarray}
As was shown in (A), the state (\ref{3-3a}) can be reexpressed as 
\begin{equation}\label{3-3b}
\ket{\lambda \kappa;ll_0}=({\hat L}_+)^{l+l_0}({\hat R}_-)^{\lambda-\kappa+l}
({\hat T}_+)^{\lambda+\kappa-l}({\hat d}_-^*)^{2l} \ket{0} \ . 
\end{equation}
\end{subequations}
Here, of course, ${\hat L}_{\pm,0}={\hat \Lambda}_{\pm,0}+{\hat K}_{\pm,0}$. 
The operators $({\hat R}_{\pm,0})$ and $({\hat T}_{\pm,0})$ are defined 
in the form 
\begin{eqnarray}
& &{\hat R}_-={\hat c}_+^*{\hat c}_- + {\hat d}_+^*{\hat d}_- \ , 
\qquad
{\hat R}_+={\hat c}_-^*{\hat c}_+ + {\hat d}_-^*{\hat d}_+ \ , \nonumber\\
& &{\hat R}_0=(1/2)({\hat c}_+^*{\hat c}_+ - {\hat c}_-^*{\hat c}_-)
+(1/2)({\hat d}_+^*{\hat d}_+ - {\hat d}_-^*{\hat d}_-) \ , 
\label{3-4}\\
& &{\hat T}_+={\hat c}_+^*{\hat d}_-^* - {\hat c}_-^*{\hat d}_+^* \ , 
\qquad
{\hat T}_-={\hat d}_-{\hat c}_+ - {\hat d}_+{\hat c}_- \ , \nonumber\\
& &{\hat T}_0=1+(1/2)({\hat c}_+^*{\hat c}_+ + {\hat d}_-^*{\hat d}_-)
+(1/2)({\hat c}_-^*{\hat c}_- + {\hat d}_+^*{\hat d}_+) \ .  
\label{3-5}
\end{eqnarray}
The operators $({\hat R}_{\pm,0})$ and $({\hat T}_{\pm,0})$ obey the 
following relations : 
\begin{subequations}\label{3-6}
\begin{eqnarray}
& &[\ {\hat R}_+ \ , \ {\hat R}_- \ ]=-2{\hat R}_0 \ , \qquad 
[\ {\hat R}_0 \ , \ {\hat R}_\pm \ ]=\mp{\hat R}_\pm \ , 
\label{3-6a}\\
& &[\ {\hat T}_+ \ , \ {\hat T}_- \ ]=-2{\hat T}_0 \ , \qquad 
[\ {\hat T}_0 \ , \ {\hat T}_\pm \ ]=\pm{\hat T}_\pm \ , 
\label{3-6b}\\
& &[\ {\rm any\ of\ }{\hat R}_{\pm,0}\ , \ 
{\rm any\ of\ }{\hat T}_{\pm,0} \ ]=0 \ . 
\label{3-6c}
\end{eqnarray}
\end{subequations}
Further, we have 
\begin{equation}\label{3-7}
[\ {\rm any\ of\ }{\hat R}_{\pm,0}\ {\rm and\ }{\hat T}_{\pm,0} \ , \ 
{\rm any\ of\ }{\hat L}_{\pm,0} \ ]=0 \ .
\end{equation}
The sets $({\hat R}_{\pm,0})$ and $({\hat T}_{\pm,0})$ form the $su(2)$- 
and the $su(1,1)$-algebra, respectively. 
We should note that the sign of the right-hand side of the commutation 
relation 
(\ref{3-6a}) is opposite of that of the relation (\ref{2-2a}). 
It appears in the quantum mechanics of rigid body. 
The Casimir operators ${\hat {\mib R}}^2$ and ${\hat {\mib T}}^2$ 
satisfy 
\begin{equation}\label{3-8}
{\hat {\mib R}}^2={\hat {\mib T}}^2={\hat {\mib L}}^2 \ . 
\end{equation}
Here, ${\hat {\mib T}}^2$ is defined as 
\begin{equation}\label{3-9}
{\hat {\mib T}}^2={\hat T}_0^2-(1/2)({\hat T}_-{\hat T}_+
+{\hat T}_+{\hat T}_-) \ . 
\end{equation}
In \S 7, the present formalism will be appeared as an application 
to the famous problem on the hydrogen atom obeying the $so(4)$-algebra.

\section{An example obeying the condition ${\hat {\mib \Gamma}}_2=0$ in 
the $so(4)$-algebra}

It may be clear that the boson operators $({\hat c}_+^*, {\hat d}_+^*)$ 
and $({\hat c}_-^*, {\hat d}_-^*)$ used in \S 3 are spherical tensors with 
rank$=1/2$ (spinors) with respect to $({\hat L}_{\pm,0})$. 
Then, the $so(4)$-algebra may be formulated in terms of the 
$su(2)\otimes su(2)$-algebra. In this section, we investigate the system 
with four kinds of bosons in terms of the boson operators which are 
spherical tensors with rank$=$1 (vector) and with 
rank$=$0 (scalar). 
We denote these operators as $({\hat c}_{\pm,0}^*, {\hat c}_{\pm,0})$ and 
$({\hat d}_0^*, {\hat d}_0)$, respectively. 
With the use of them, ${\hat L}_{\pm,0}$ and ${\hat M}_{\pm,0}$ can be 
expressed as 
\begin{subequations}\label{4-1}
\begin{eqnarray}
& &{\hat L}_+=\sqrt{2}({\hat c}_+^*{\hat c}_0 + {\hat c}_0^*{\hat c}_-) \ , 
\nonumber\\
& &{\hat L}_-=\sqrt{2}({\hat c}_0^*{\hat c}_+ + {\hat c}_-^*{\hat c}_0) \ , 
\nonumber\\
& &{\hat L}_0={\hat c}_+^*{\hat c}_+ - {\hat c}_-^*{\hat c}_- \ , 
\label{4-1a}\\
& &{\hat M}_+=\sqrt{2}({\hat c}_+^*{\hat d}_0 - {\hat d}_0^*{\hat c}_-) \ , 
\nonumber\\
& &{\hat M}_-=\sqrt{2}({\hat d}_0^*{\hat c}_+ - {\hat c}_-^*{\hat d}_0) \ , 
\nonumber\\
& &{\hat M}_0=-{\hat c}_0^*{\hat d}_0 - {\hat d}_0^*{\hat c}_0 \ . 
\label{4-1b}
\end{eqnarray}
\end{subequations}
They obey the $so(4)$-algebra and straightforward calculation gives us 
identically 
\begin{equation}\label{4-2}
{\hat {\mib \Gamma}}_2=(1/2){\hat {\mib L}}\cdot{\hat {\mib M}}=0 \ , 
\quad {\rm i.e.,}\quad 
{\hat {\mib \Gamma}}_1={\hat {\mib \Lambda}}^2={\hat {\mib K}}^2 \ . 
\end{equation}

For the sets $({\hat L}_{\pm,0},{\hat M}_{\pm,0})$, let us investigate the 
orthogonal set. 
In \S 2, we mentioned that, in addition to the set 
$({\hat {\mib \Gamma}}_1, {\hat \Lambda}_0, {\hat K}_0)$ or 
$({\hat {\mib \Gamma}}_1 , {\hat {\mib L}}^2 , {\hat L}_0)$, we must find 
one more hermitian operator. 
For this purpose, we introduce the following operators appearing in 
the relation (B$\cdot$5$\cdot$1a) : 
\begin{eqnarray}\label{4-3}
& &{\hat H}_+={\hat c}_+^*{\hat c}_-^*-(1/2){\hat c}_0^{*2} +
(1/2){\hat d}_0^{*2} \ , \nonumber\\
& &{\hat H}_-={\hat c}_-{\hat c}_+-(1/2){\hat c}_0^{2} +
(1/2){\hat d}_0^{2} \ , \nonumber\\
& &{\hat H}_0=1+(1/2)({\hat c}_+^*{\hat c}_+ +{\hat c}_0^{*}{\hat c}_0 +
{\hat c}_-^*{\hat c}_-)+(1/2){\hat d}_0^{*}{\hat d}_0 \ . 
\end{eqnarray}
The set $({\hat H}_{\pm,0})$ obeys the $su(1,1)$-algebra : 
\begin{equation}\label{4-4}
[\ {\hat H}_+\ , \ {\hat H}_- \ ]=-2{\hat H}_0 \ , \qquad
[\ {\hat H}_0\ , \ {\hat H}_\pm \ ]=\pm{\hat H}_\pm \ .
\end{equation}
The Casimir operator ${\hat {\mib H}}^2$, which is of the same form as 
${\hat {\mib T}}^2$ defined in the relation (\ref{3-9}), is identically equal 
to 
\begin{equation}\label{4-5}
{\hat {\mib H}}^2={\hat {\mib \Gamma}}_1={\hat {\mib \Lambda}}^2
={\hat {\mib K}}^2 \ . 
\end{equation}
Further, we have 
\begin{equation}\label{4-6}
[\ {\rm any\ of\ }({\hat H}_{\pm,0})\ , \ {\rm any\ of\ }
({\hat \Lambda}_{\pm,0}\ ,\ {\hat K}_{\pm,0}) \ ]=0 \ . 
\end{equation}
Then, as the operator additional to $({\hat {\mib \Gamma}}_1,{\hat \Lambda}_0, 
{\hat K}_0)$, we can choose ${\hat H}_0$, i.e., \break
$({\hat {\mib \Gamma}}_1, {\hat \Lambda}_0, {\hat K}_0, {\hat H}_0)$.

We are now possible to find the eigenstates for the set 
$({\hat {\mib \Gamma}}_1, {\hat \Lambda}_0, {\hat K}_0, {\hat H}_0)$. 
First, we note that the algebraic structure of the set 
$({\hat \Lambda}_{\pm,0}, {\hat K}_{\pm,0}, {\hat H}_{\pm,0})$ is 
essentially the same as that of the set 
$({\hat L}_{\pm,0}, {\hat R}_{\mp,0}, {\hat T}_{\pm,0})$ used in \S 3.  
Then, we are able to derive the following state : 
\begin{eqnarray}\label{4-7}
& &\ket{\mu;\lambda_0 \kappa_0 \eta_0}=({\hat \Lambda}_+)^{\mu+\lambda_0}
({\hat K}_+)^{\mu+\kappa_0}({\hat H}_+)^{\eta_0-(\mu+1)}({\hat c}_-^*)^{2\mu}
\ket{0} \ , \nonumber\\
& &\qquad
\mu=0,\ 1/2,\ 1,\ 3/2,\cdots , \nonumber\\
& &\qquad \lambda_0,\ \kappa_0 =-\mu,\ -\mu+1, \ \cdots, \ \mu-1,\ \mu \ , 
\nonumber\\
& &\qquad \eta_0=\mu+1,\ \mu+2,\ \mu+3,\cdots \ . 
\end{eqnarray}
The eigenvalue of ${\hat {\mib \Gamma}}_1$ is given by $\mu(\mu+1)$. 
With the use of the Clebsch-Gordan coefficient, we obtain the eigenstate 
of the set 
$({\hat {\mib \Gamma}}_1, {\hat H}_0, {\hat {\mib L}}^2, {\hat L}_0)$ : 
\begin{eqnarray}\label{4-8}
& &\ket{\mu\eta_0;ll_0}=\sum_{\lambda_0,\kappa_0}
\bra{\mu\lambda_0,\mu\kappa_0}ll_0\rangle \ket{\mu;\lambda_0\kappa_0\eta_0} \ ,
\nonumber\\
& &\qquad l=0,\ 1,\ 2,\cdots\ , \qquad
l_0=-l,\ -l+1,\ \cdots l-1,\ l \ . 
\end{eqnarray}
In the present case, $l$ is restricted to integer. 
It may be understandable from the fact that the set $({\hat L}_{\pm,0})$ 
is composed of the boson $({\hat c}_{\pm,0}^*,{\hat c}_{\pm,0})$ 
with rank$=1$.

Our final task of this section is to present a possible expression of 
$\ket{\mu\eta_0;ll_0}$ in the form of a monomial such as 
the state (\ref{3-3b}). 
Let the state (\ref{4-8}) express in the form 
\begin{equation}\label{4-9}
\ket{\mu\eta_0;ll_0}=({\hat L}_+)^{l+l_0}\ket{\mu\eta_0;l} \ .
\end{equation}
Here, $\ket{\mu\eta_0;l}$ should satisfy 
\begin{eqnarray}
& &{\hat L}_-\ket{\mu\eta_0;l}=0 \ , \qquad 
{\hat L}_0\ket{\mu\eta_0;l}=-l\ket{\mu\eta_0;l}\ , 
\label{4-10}\\
& &{\hat {\mib H}}^2\ket{\mu\eta_0;l}=(\mu+1)[(\mu+1)-1]\ket{\mu\eta_0;l} \ , 
\quad ({\hat {\mib H}}^2={\hat {\mib \Gamma}}_1) \nonumber\\
& &{\hat H}_0\ket{\mu\eta_0;l}=\eta_0\ket{\mu\eta_0;l}\ . 
\label{4-11}
\end{eqnarray}
For obtaining the state $\ket{\mu\eta_0;l}$, we note the state 
(B$\cdot$5$\cdot$19a). 
Let ${\hat T}_+$, $t$, $t_0$ and $w$ in this state read 
${\hat H}_+$, $\mu+1$, $\eta_0$ and $(l+1)/2$, respectively, in the present 
notations. 
Then, the state $\ket{\mu\eta_0;l}$ satisfying the condition (\ref{4-10}) 
and (\ref{4-11}) are given in the following :\break 
\begin{eqnarray}\label{4-12}
\ket{\mu\eta_0;l}&=&
({\hat H}_+)^{\eta_0-\mu-1}\left[
{\hat X}_+({\hat X}_0+l/2+3/4+\varepsilon)^{-1}
-{\hat Y}_+({\hat Y}_0+y+\varepsilon)^{-1}\right]^{\mu+1/4-l/2-y} \nonumber\\
& &\times \ket{(l+1)/2\ (-), 1/4, y}\ . \qquad (l=0,\ 1,\ 2,\cdots)
\end{eqnarray}
The proof is given by noting that any of $({\hat L}_{\pm,0})$ commutes with 
any of $({\hat X}_{\pm,0}$ and ${\hat Y}_{\pm,0})$ defined in the 
relations (B$\cdot$5$\cdot$2) and (B$\cdot$5$\cdot$3) and the relation 
\begin{equation}\label{4-13}
{\hat L}_-\ket{(l+1)/2,1/4,y}=0 \ , \qquad
{\hat L}_0\ket{(l+1)/2,1/4,y}=-l\ket{(l+1)/2,1/4,y} \ . 
\end{equation}
It may be self-evident from the discussion in (B) that 
$\ket{\mu\eta_0;l}$ is the eigenstate of ${\hat {\mib H}}^2$ and ${\hat H}_0$ 
with the eigenvalues $(\mu+1)[(\mu+1)-1]$ and $\eta_0$, respectively. 
Since $y=1/4$ and $3/4$, the state $\ket{(l+1)/2\ (-), 1/4, y}$ can be 
written as 
\begin{eqnarray}\label{4-14}
& &\ket{(l+1)/2\ (-),1/4,1/4}=({\hat c}_-^*)^l \ket{0}\ , \nonumber\\
& &\ket{(l+1)/2\ (-),1/4,3/4}=({\hat c}_-^*)^l {\hat d}_0^*\ket{0}\ .
\end{eqnarray}
Since $\eta_0-\mu-1=0,\ 1,\ 2,\cdots$, we have 
\begin{equation}\label{4-15}
\eta_0=\mu+1,\ \mu+2,\ \mu+3,\cdots \ . 
\end{equation}
Further, $\mu+1/4-l/2-y=0,\ 1,\ 2, \cdots$ and 
for $y=1/4$ and $3/4$, the following rule is derived : 
\begin{equation}\label{4-16}
\mu=\biggl\{\begin{array}{ll}
l/2,\ l/2+1,\ l/2+2, \cdots \ , & (y=1/4) \\ 
l/2+1/2,\ l/2+3/2,\ l/2+5/2, \cdots \ , & (y=3/4) 
\end{array}
\end{equation}
If the eigenvalue of ${\hat {\mib H}}^2$ is denoted as $\eta(\eta-1)$, 
the rule 
(\ref{4-16}) is rewritten as 
\begin{equation}\label{4-17}
\eta=\biggl\{\begin{array}{ll}
l/2+1,\ l/2+2, \ l/2+3, \cdots \ , & (y=1/4) \\ 
l/2+3/2,\ l/2+5/2,\ l/2+7/2, \cdots \ , & (y=3/4) 
\end{array}
\end{equation}
Therefore, $y$ plays a role of classifying the relation between $\mu$ 
or $\eta$ and $l$ into two groups.

\section{An example obeying the condition ${\wtilde {\mib \Gamma}}_2\neq 0$ in 
the $so(3,1)$-algebra}

Main aim of this section is to present a concrete example of the 
$so(3,1)$-algebra under the condition ${\wtilde {\mib \Gamma}}_2\neq 0$. 
We formulate this case in terms of the boson operators 
$({\hat c}_{\pm}^*, {\hat c}_{\pm})$ and $({\hat d}_{\pm}^*, {\hat d}_{\pm})$ 
used in \S 3. 
The formalism is in parallel to that presented in \S 3. 
With the use of these bosons, the sets $({\hat L}_{\pm,0})$ and 
$({\wtilde M}_{\pm,0})$ are defined in the form 
\begin{subequations}\label{5-1}
\begin{eqnarray}
& &{\hat L}_+={\hat c}_+^*{\hat d}_+ + {\hat c}_-^*{\hat d}_- \ , 
\qquad
{\hat L}_-={\hat d}_+^*{\hat c}_+ + {\hat d}_-^*{\hat c}_- \ , 
\nonumber\\
& &{\hat L}_0=(1/2)({\hat c}_+^*{\hat c}_+ - {\hat d}_+^*{\hat d}_+)
+ (1/2)({\hat c}_-^*{\hat c}_- - {\hat d}_-^*{\hat d}_-) \ , 
\label{5-1a}\\
& &{\wtilde M}_+=-{\hat c}_+^*{\hat c}_-^* + {\hat d}_-{\hat d}_+ \ , 
\qquad
{\wtilde M}_-=-{\hat c}_-{\hat c}_+ + {\hat d}_+^*{\hat d}_-^* \ , 
\nonumber\\
& &{\wtilde M}_0=(1/2)({\hat c}_+^*{\hat d}_-^* + {\hat d}_-{\hat c}_+)
+ (1/2)({\hat c}_-^*{\hat d}_+^* + {\hat d}_+{\hat c}_-) \ . 
\label{5-1b}
\end{eqnarray}
\end{subequations}
The set $({\hat L}_{\pm,0})$ is of the same form as that used in \S 3 and 
straightforward calculation gives us the commutation relations for the 
$so(3,1)$-algebra. 
The Casimir operators ${\wtilde {\mib \Gamma}}_1$ ($=(1/4)({\hat {\mib L}}^2
-{\wtilde {\mib M}}^2)$) and 
${\wtilde {\mib \Gamma}}_2$ ($=(1/2){\hat {\mib L}}\cdot{\wtilde {\mib M}}$) 
are expressed as follows : 
\begin{subequations}\label{5-2}
\begin{eqnarray}
& &{\wtilde {\mib \Gamma}}_1=(1/4)(-1+{\wtilde F}^2-{\wtilde G}^2) \ , 
\label{5-2a}\\
& &{\wtilde {\mib \Gamma}}_2=-(1/2){\wtilde F}{\wtilde G} \ . 
\label{5-2b}
\end{eqnarray}
\end{subequations}
Here, ${\wtilde F}$ and ${\wtilde G}$ are defined as 
\begin{subequations}\label{5-3}
\begin{eqnarray}
& &{\wtilde F}=(1/2)({\hat c}_+^*{\hat c}_+ - {\hat d}_-^*{\hat d}_-) 
-(1/2)({\hat c}_-^*{\hat c}_- - {\hat d}_+^*{\hat d}_+) \ , 
\label{5-3a}\\
& &{\wtilde G}=(1/2)({\hat c}_+^*{\hat d}_-^* + {\hat d}_-{\hat c}_+) 
-(1/2)({\hat c}_-^*{\hat d}_+^* + {\hat d}_+{\hat c}_-) \ . 
\label{5-3b}
\end{eqnarray}
\end{subequations}
It should be noted that any of $({\hat L}_{\pm,0}, {\wtilde M}_{\pm,0})$ 
commutes with ${\wtilde F}$ and ${\wtilde G}$ which are mutually commutable.

As for the mutually commutable operators, we can adopt ${\wtilde F}$, 
${\wtilde G}$, ${\hat L}_0$ and ${\wtilde M}_0$, which are 
equivalent to ${\wtilde {\mib \Gamma}}_1$, 
${\wtilde {\mib \Gamma}}_2$, ${\hat L}_0$ and ${\wtilde M}_0$. 
Then, in terms of the linear combinations of these operators, 
four mutually commutable operators $(1/2)({\wtilde M}_0 \pm {\wtilde G})$ 
and $(1/2)({\hat L}_0\pm{\wtilde F}+1)$ can be introduced : 
\begin{eqnarray}
& &(1/2)({\wtilde M}_0 \pm {\wtilde G})=(1/2)({\hat c}_{\pm}^*{\hat d}_{\mp}^*
+{\hat d}_{\mp}{\hat c}_{\pm}) \ , 
\label{5-4}\\ 
& &(1/2)({\hat L}_0\pm{\wtilde F}+1)=1/2+
(1/2)({\hat c}_{\pm}^*{\hat c}_{\pm}
-{\hat d}_{\mp}^*{\hat d}_{\mp}) \ . 
\label{5-5} 
\end{eqnarray}
The forms (\ref{5-4}) and (\ref{5-5}) permit us to introduce two independent 
$su(1,1)$-algebra : 
\begin{subequations}\label{5-6}
\begin{eqnarray}
& &{\wtilde \Lambda}_+={\hat c}_+^*{\hat d}_-^* \ , \quad
{\wtilde \Lambda}_-={\hat d}_-{\hat c}_+ \ , \quad
{\wtilde \Lambda}_0=1/2+(1/2)({\hat c}_+^*{\hat c}_+ + 
{\hat d}_-^*{\hat d}_-) \ , 
\label{5-6a}\\
& &{\wtilde K}_+={\hat c}_-^*{\hat d}_+^* \ , \quad
{\wtilde K}_-={\hat d}_+{\hat c}_- \ , \quad
{\wtilde K}_0=1/2+(1/2)({\hat c}_-^*{\hat c}_- + 
{\hat d}_+^*{\hat d}_+) \ . 
\label{5-6b}
\end{eqnarray}
\end{subequations}
The above are in parallel to the forms (\ref{3-1a}) and (\ref{3-1b}) for the 
$su(1,1)$-algebras. 
The operators introduced in the relation (\ref{5-4}) can be expressed as 
\begin{subequations}\label{5-7}
\begin{eqnarray}
& &(1/2)({\wtilde M}_0 + {\wtilde G})
=(1/2)({\wtilde \Lambda}_+ + {\wtilde \Lambda}_-)\ (={\wtilde \Lambda}^0) \ , 
\label{5-7a}\\
& &(1/2)({\wtilde M}_0 - {\wtilde G})
=(1/2)({\wtilde K}_+ + {\wtilde K}_-)\ (={\wtilde K}^0) \ . 
\label{5-7b}
\end{eqnarray}
\end{subequations}
Further, we denote the operators introduced in the relation (\ref{5-5}) as 
${\wtilde \Lambda}$ and ${\wtilde K}$ : 
\begin{subequations}\label{5-8}
\begin{eqnarray}
& &(1/2)({\hat L}_0 + {\wtilde F}+1)={\wtilde \Lambda}
=1/2+(1/2)({\hat c}_+^*{\hat c}_+ - {\hat d}_-^*{\hat d}_-) \ , 
\label{5-8a}\\
& &(1/2)({\hat L}_0 - {\wtilde F}+1)={\wtilde K}
=1/2+(1/2)({\hat c}_-^*{\hat c}_- - {\hat d}_+^*{\hat d}_+) \ . 
\label{5-8b}
\end{eqnarray}
\end{subequations}
The Casimir operators of the $su(1,1)$-algebras $({\wtilde \Lambda}_{\pm,0})$ 
and $({\wtilde K}_{\pm,0})$ can be expressed as 
\begin{subequations}\label{5-9}
\begin{eqnarray}
& &{\wtilde {\mib \Lambda}}^2={\wtilde \Lambda}({\wtilde \Lambda}-1) \ , 
\label{5-9a}\\
& &{\wtilde {\mib K}}^2={\wtilde K}({\wtilde K}-1) \ . 
\label{5-9b}
\end{eqnarray}
\end{subequations}
From the above argument, we can conclude that, in the present framework, the 
$su(1,1)\otimes su(1,1)$-algebra is constructed. 
Mutually commutable four operators in this algebra are identical to 
those of the original $so(3,1)$-algebra presented in the relation (\ref{5-1}), 
i.e., $({\wtilde {\mib \Lambda}}^2, {\wtilde \Lambda}^0)$ and 
$({\wtilde {\mib K}}^2, {\wtilde K}^0)$.

Concerning the eigenvalue problem for 
$({\wtilde {\mib \Lambda}}^2, {\wtilde \Lambda}^0)$ and 
$({\wtilde {\mib K}}^2, {\wtilde K}^0)$, we presented the basic idea in (B). 
It can be seen in various relations in \S\S 2 and 3 of (B). 
In this section, 
we summarize the results given in \S 3 of (B) 
in terms of the form suitable to the present 
system. 
We denote the eigenvalues of ${\wtilde {\mib \Lambda}}^2$, 
${\wtilde \Lambda}^0$, ${\wtilde {\mib K}}^2$ and ${\wtilde K}^0$ as 
$\lambda(\lambda-1)$, $i\rho\lambda^0$ $(\rho=\pm)$, $\kappa(\kappa-1)$ 
and $i\sigma\kappa^0$ $(\sigma=\pm)$, respectively. 
Then, the eigenstate, which is denoted as $\kket{(\rho u)\lambda\lambda^0;
(\sigma v)\kappa\kappa^0}$ is expressed in the following form :
\begin{eqnarray}\label{5-10}
& &\kket{(\rho u)\lambda\lambda^0; (\sigma v)\kappa\kappa^0}
=({\wtilde \Lambda}^{\rho})^{\lambda^0-\lambda}\kket{\rho\lambda(u)}
\otimes ({\wtilde K}^{\sigma})^{\kappa^0-\kappa}\kket{\sigma\kappa (v)} \ . 
\end{eqnarray}
Here, $\kket{\rho\lambda (u)}$ and $\kket{\sigma\kappa (v)}$ are given by 
\begin{subequations}\label{5-11}
\begin{eqnarray}
& &\kket{\rho\lambda (u)}=\exp(i\rho{\wtilde \Lambda}^+)\ket{\lambda (u)}\ , 
\qquad (\rho=\pm) 
\label{5-11a}\\
& &\kket{\sigma\kappa (v)}=\exp(i\sigma{\wtilde K}^+)\ket{\kappa (v)}\ , 
\qquad (\sigma=\pm) \qquad\qquad\qquad\quad
\label{5-11b}
\end{eqnarray}
\end{subequations}
\vspace{-0.4cm}
\begin{subequations}\label{5-12}
\begin{eqnarray}
& &\ket{\lambda (u)}=\biggl\{\begin{array}{ll}
({\hat c}_+^*)^{2\lambda-1}\ket{0} \ , 
& (u=+,\ \lambda=1,\ 3/2,\ 2, \cdots) 
\\
({\hat c}_-^*)^{2\lambda-1}\ket{0} \ , & (u=-,\ \lambda=1/2,\ 1,\ 3/2, \cdots)
\end{array}
\label{5-12a}\\
& &\ket{\kappa (v)}=\biggl\{\begin{array}{ll}
({\hat d}_+^*)^{2\kappa-1}\ket{0} \ , & (v=+,\ \kappa=1,\ 3/2,\ 2, \cdots) 
\\
({\hat d}_-^*)^{2\kappa-1}\ket{0} \ . & (v=-,\ \kappa=1/2,\ 1,\ 3/2, \cdots)
\end{array}
\label{5-12b}
\end{eqnarray}
\end{subequations}
The operators ${\wtilde \Lambda}^\rho$ and ${\wtilde K}^{\sigma}$ are defined in the form 
\begin{equation}\label{5-13}
{\wtilde \Lambda}^\rho=(1/2i)({\wtilde \Lambda}_+ - {\wtilde \Lambda}_-)
-\rho{\wtilde \Lambda}_0 \ , \qquad
{\wtilde K}^\sigma=(1/2i)({\wtilde K}_+ - {\wtilde K}_-)
-\sigma{\wtilde K}_0 \ .
\end{equation}
The normalization constant for any state is omitted.

Our next concern is to investigate the eigenvalue problem for 
${\wtilde F}$, ${\wtilde G}$, ${\hat {\mib L}}^2$ and ${\hat L}_0$. 
First, we note the following relations : 
\begin{eqnarray}
& &{\hat T}_{\pm}={\wtilde \Lambda}_{\pm}-{\wtilde K}_{\pm} \ , \qquad
{\hat T}_0={\wtilde \Lambda}_0+{\wtilde K}_0 \ , 
\label{5-14}\\
& &{\wtilde F}={\hat R}_0 \ , \qquad {\wtilde G}=(1/2)({\hat T}_+ 
+ {\hat T}_-)\ (={\hat T}^0) \ . 
\label{5-15}
\end{eqnarray}
Here, $({\hat T}_{\pm,0})$ was already given in the relation (\ref{3-5}) 
and ${\hat R}_0$ was also defined as a member of the $su(2)$-algebra 
$({\hat R}_{\pm,0})$ shown in the relation (\ref{3-4}). 
Therefore, we are able to see that the present system is essentially 
equivalent to that discussed in the second half of \S 3. 
Combining this discussion with that given in (B), we obtain the eigenstate of 
${\wtilde F}\ (={\hat R}_0)$, ${\wtilde G}\ (={\hat T}^0)$, 
${\hat {\mib L}}^2$ 
and ${\hat L}_0$ with the eigenvalues $r_0$, $\pm i t^0$, $l(l+1)$ and $l_0$ 
in the following form : 
\begin{eqnarray}\label{5-16}
\kket{r_0,\pm t^0;ll_0}&=&
({\hat T}^{\pm})^{t^0-(l+1)}\exp(\pm i{\hat T}_+)
({\hat L}_+)^{l+l_0}({\hat R}_-)^{l+r_0}({\hat d}_-^*)^{2l}\ket{0}
\nonumber\\
&=&({\hat L}_+)^{l+l_0}({\hat T}^{\pm})^{t^0-(l+1)}\exp(\pm i {\hat T}_+)
({\hat R}_-)^{l+r_0}({\hat d}_-^*)^{2l}\ket{0}\ . 
\end{eqnarray}
Here, it should be noted that the eigenvalues of ${\hat {\mib T}}^2$, 
${\hat {\mib R}}^2$ and ${\hat {\mib L}}^2$ are equal to $l(l+1)$.

\section{An example obeying the condition ${\wtilde {\mib \Gamma}}_2=0$ 
in the $so(3,1)$-algebra}

Our final problem is to investigate a possible example obeying the 
condition ${\wtilde {\mib \Gamma}}_2=0$ in the $so(3,1)$-algebra. 
We use the boson operators $({\hat c}_{\pm,0}^*, {\hat c}_{\pm,0})$ and 
$({\hat d}_0^*, {\hat d}_0)$ adopted in \S 4. 
With the use of these bosons, we define the following operators : 
\begin{eqnarray}\label{6-1}
& &{\wtilde M}_+=\sqrt{2}(-{\hat c}_+^*{\hat d}_0^* + {\hat d}_0{\hat c}_-)\ , 
\nonumber\\
& &{\wtilde M}_-=\sqrt{2}(-{\hat d}_0{\hat c}_+ + {\hat c}_-^*{\hat d}_0^*)\ , 
\nonumber\\
& &{\wtilde M}_0={\hat c}_0^*{\hat d}_0^* + {\hat d}_0{\hat c}_0 \ . 
\end{eqnarray}
The set $({\hat L}_{\pm,0})$ defined in the relation (\ref{4-1a}) and the set 
$({\wtilde M}_{\pm,0})$ defined in the relation (\ref{6-1}) compose the 
$so(3,1)$-algebra and it is in parallel to the $so(4)$-algebra in the 
representation (\ref{4-1}). 
Straightforward calculation of ${\wtilde {\mib \Gamma}}_2$ defined in the 
relation (\ref{2-12}), in the present case, leads us identically 
to 
\begin{equation}\label{6-2}
{\wtilde {\mib \Gamma}}_2=(1/2){\hat {\mib L}}\cdot{\wtilde {\mib M}}=0 \ .
\end{equation}
In the case of ${\wtilde {\mib \Gamma}}_1$, we have the following form : 
\begin{equation}\label{6-3}
{\wtilde {\mib \Gamma}}_1=(1/4)({\hat {\mib L}}^2-
{\wtilde {\mib M}}^2)={\wtilde {\mib H}}^2 \ .
\end{equation}
Here, ${\wtilde {\mib H}}^2$ is the Casimir operator of the 
$su(1,1)$-algebra $({\wtilde H}_{\pm,0})$ defined as 
\begin{eqnarray}\label{6-4}
& &{\wtilde H}_+={\hat c}_+^*{\hat c}_-^* - (1/2){\hat c}_0^{*2}
-(1/2){\hat d}_0^2 \ , \nonumber\\
& &{\wtilde H}_-={\hat c}_-{\hat c}_+ - (1/2){\hat c}_0^{2}
-(1/2){\hat d}_0^{*2} \ , \nonumber\\
& &{\wtilde H}_0=1/2+(1/2)({\hat c}_+^*{\hat c}_+ + {\hat c}_0^*{\hat c}_0 
+ {\hat c}_-^* {\hat c}_-) - (1/2){\hat d}_0^{*}{\hat d}_0 \ . 
\end{eqnarray}
The representation (\ref{6-4}) was already introduced in the relation 
(B$\cdot$5$\cdot$1b) under the notation $({\wtilde T}_{\pm,0})$. 
It may be interesting to see the relation 
\begin{equation}\label{6-5}
[\ {\rm any\ of\ }({\wtilde H}_{\pm,0})\ , \ {\rm any\ of\ }
({\hat L}_{\pm,0}\ , \ {\wtilde M}_{\pm,0})\ ] =0 \ .
\end{equation}
Then, we can search the eigenstate of 
$({\wtilde {\mib \Gamma}}_1={\wtilde {\mib H}}^2, 
{\wtilde H}_0, {\hat {\mib L}}^2, {\hat L}_0)$.

Let the eigenvalues of ${\wtilde {\mib H}}^2$, ${\wtilde H}_0$, 
${\hat {\mib L}}^2$ and ${\hat L}_0$ denote as $\eta(\eta-1)$, $\eta_0$, 
$l(l+1)$ and $l_0$, respectively. 
Its eigenstate $\rket{\eta\eta_0;ll_0}$ can be expressed as follows : 
\begin{equation}\label{6-6}
\rket{\eta\eta_0;ll_0}=({\hat L}_+)^{l+l_0}\rket{\eta\eta_0;l}\ . 
\end{equation}
Here, $\rket{\eta\eta_0;l}$ should satisfy 
\begin{eqnarray}
& &{\hat L}_-\rket{\eta\eta_0;l}=0 \ , \qquad
{\hat L}_0\rket{\eta\eta_0;l}=-l\rket{\eta\eta_0;l} \ , 
\label{6-7}\\
& &{\wtilde {\mib H}}^2\rket{\eta\eta_0;l}=\eta(\eta-1)\rket{\eta\eta_0;l} \ ,
\qquad ({\wtilde {\mib H}}^2={\wtilde {\mib \Gamma}}_1) \nonumber\\
& &{\wtilde H}_0\rket{\eta\eta_0;l}=\eta_0\rket{\eta\eta_0;l} \ . 
\label{6-8}
\end{eqnarray}
For obtaining the state $\rket{\eta\eta_0;ll_0}$, we note the state 
(B$\cdot$5$\cdot$19b). 
If ${\wtilde T}_+$, $t$, $t_0$ and $w$ in this state read ${\wtilde H}_+$, 
$\eta$, $\eta_0$ and $(l+1)/2$, respectively, in the present notations, 
the state $\rket{\eta\eta_0;l}$ obeying the conditions (\ref{6-7}) and 
(\ref{6-8}) can be expressed as 
\begin{eqnarray}\label{6-9}
\rket{\eta\eta_0;l}&=&
({\wtilde H}_+)^{\eta_0-\eta}({\hat Y}_+)^{l/2+3/4-y-\eta}\nonumber\\
& &\times
\exp\left[{\hat X}_+({\hat X}_0+l/2+3/4+\varepsilon)^{-1}{\hat Y}_+\right]
\ket{(l+1)/2\ (-),1/4,y} \ .\quad
\end{eqnarray}
Here, $\ket{(l+1)/2\ (-),1/4,y}$ is defined in the relation (\ref{4-14}). 
Of course, $\eta_0$ satisfies 
\begin{equation}\label{6-10}
\eta_0=\eta,\ \eta+1,\ \eta+2, \cdots \ . 
\end{equation}
Further, $0\leq l/2+3/4-y-\eta=0,\ 1,\ 2,\ \cdots$ and 
for $y=1/4$ and $3/4$, the following rule is obtained : 
\begin{equation}\label{6-11}
\eta=\Biggl\{\begin{array}{ll}
l/2+1/2,\ l/2-1/2,\ \cdots &
\biggl\{\begin{array}{ll}
1/2 & (l : even)  \\ 1 & (l : {\rm odd})  \ , \ \  (y=1/4) 
\end{array}\\
l/2,\ l/2-1,\ \cdots &
\biggl\{\begin{array}{ll}
1 & (l : {\rm even})  \\ 1/2 & (l : {\rm odd})  \ .\ \  (y=3/4) 
\end{array}
\end{array}
\end{equation}
The rule (\ref{6-11}) should be compared with the rule (\ref{4-17}). 
It should be also noted that the relation (B$\cdot$4$\cdot$8), under which the 
state (\ref{6-9}) is normalizable, gives us 
\begin{equation}\label{6-12}
l>0\ , \quad ({\rm for}\ y=1/4)\ , \qquad 
l>1\ , ({\rm for}\ y=3/4) \ .
\end{equation}

\section{A possible application}

As was mentioned in \S 1, the main aim of investigating the $so(4)$- 
and the $so(3,1)$-algebra in the boson realization 
is to apply its idea to the $so(4)$-model for describing nuclear dynamics. 
In the forthcoming paper, we will discuss this problem. 
In this section, we investigate the $so(4)$- and the $so(3,1)$-algebra 
presented in \S 3 and \S 5, respectively, in 
more detail, especially, in relation to the hydrogen atom and the 
scattering problem.

Our discussion starts in the following simple Schr\"odinger equation : 
\begin{eqnarray}
{\hat {\cal H}}_4\ket{\Psi}&=&{\cal E}\ket{\Psi} \ , 
\label{7-1}\\
{\hat {\cal H}}_4&=&\omega({\hat c}_+^*{\hat c}_+ + {\hat d}_+^*{\hat d}_+ 
+{\hat c}_-^*{\hat c}_- + {\hat d}_-^*{\hat d}_- +2)\nonumber\\
&=&2\omega{\hat T}_0 \ . 
\label{7-2}
\end{eqnarray}
Here, $\omega$ represents the constant. Of course, the system under 
investigation is harmonic oscillator in four dimensional space with the 
frequency $\omega$.
The boson operators ${\hat c}_{\pm}^*$, ${\hat c}_{\pm}$, ${\hat d}_{\pm}^*$ 
and ${\hat d}_\pm$ can be parameterized in terms of the variables 
$(x_+, y_+,x_-,y_-)$ in the rectangular coordinate system : 
\begin{subequations}\label{7-3}
\begin{eqnarray}
& &{\hat c}_+^*=(1/2)\left[\sqrt{\omega}(x_+ + iy_+)-(\sqrt{\omega})^{-1}
(\partial_{x_+}+i\partial_{y_+})\right] \ , 
\nonumber\\
& &{\hat c}_+=(1/2)\left[\sqrt{\omega}(x_+ - iy_+)+(\sqrt{\omega})^{-1}
(\partial_{x_+}-i\partial_{y_+})\right] \ , 
\label{7-3a}\\
& &{\hat c}_-^*=(1/2)\left[\sqrt{\omega}(x_- + iy_-)-(\sqrt{\omega})^{-1}
(\partial_{x_-}+i\partial_{y_-})\right] \ , 
\nonumber\\
& &{\hat c}_-=(1/2)\left[\sqrt{\omega}(x_- - iy_-)+(\sqrt{\omega})^{-1}
(\partial_{x_-}-i\partial_{y_-})\right] \ , 
\label{7-3b}\\
& &{\hat d}_+^*=(1/2)\left[\sqrt{\omega}(x_- - iy_-)-(\sqrt{\omega})^{-1}
(\partial_{x_-}-i\partial_{y_-})\right] \ , 
\nonumber\\
& &{\hat d}_+=(1/2)\left[\sqrt{\omega}(x_- + iy_-)+(\sqrt{\omega})^{-1}
(\partial_{x_-}+i\partial_{y_-})\right] \ , 
\label{7-3c}\\
& &{\hat d}_-^*=-(1/2)\left[\sqrt{\omega}(x_+ - iy_+)-(\sqrt{\omega})^{-1}
(\partial_{x_+}-i\partial_{y_+})\right] \ , 
\nonumber\\
& &{\hat d}_-=-(1/2)\left[\sqrt{\omega}(x_+ + iy_+)+(\sqrt{\omega})^{-1}
(\partial_{x_+}+i\partial_{y_+})\right] \ . 
\label{7-3d}
\end{eqnarray}
\end{subequations}
For the integration, the volume element $dV_4$ is given by 
\begin{equation}\label{7-4}
dV_4=dx_+dy_+dx_-dy_- \ .
\end{equation}
The Hamiltonian (\ref{7-2}) can be rewritten as 
\begin{eqnarray}\label{7-5}
{\hat {\cal H}}_4&=&-(1/2)(\partial_{x_+}^2+\partial_{y_+}^2+\partial_{x_-}^2+
\partial_{y_-}^2) 
+(1/2)\omega^2(x_+^2+y_+^2+x_-^2+y_-^2) \ . \quad
\end{eqnarray}

The six operators composing the $so(4)$-algebra are expressed as follows : 
\begin{eqnarray}
& &{\hat L}_x=(i/2)[-y_-\partial_{x_+}-x_-\partial_{y_+}
+y_+\partial_{x_-}+x_+\partial_{y_-}] \ , \nonumber\\
& &{\hat L}_y=(i/2)[x_-\partial_{x_+}-y_-\partial_{y_+}
-x_+\partial_{x_-}+y_+\partial_{y_-}] \ , \nonumber\\
& &{\hat L}_z=(i/2)[y_+\partial_{x_+}-x_+\partial_{y_+}
+y_-\partial_{x_-}-x_-\partial_{y_-}] \ , 
\label{7-6}\\
& &{\hat M}_x=(1/2)[\omega(x_+x_- - y_+y_-)-\omega^{-1}
(\partial_{x_+}\partial_{x_-}-\partial_{y_+}\partial_{y_-})] \ , \nonumber\\
& &{\hat M}_y=(1/2)[\omega(x_+y_- - y_+x_-)-\omega^{-1}
(\partial_{x_+}\partial_{y_-}-\partial_{y_+}\partial_{x_-})] \ , \nonumber\\
& &{\hat M}_z=(1/4)[\omega(x_+^2+y_+^2-x_-^2-y_-^2)
-\omega^{-1}(\partial_{x_+}^2+\partial_{y_+}^2-\partial_{x_-}^2
-\partial_{y_-}^2)] \ . 
\label{7-7}
\end{eqnarray}
Here, for $(A_{\pm,0})$, ${\mib A}=(A_x, A_y,A_z)$ is defined as 
\begin{equation}\label{7-8}
A_x=(1/2)(A_++A_-)\ , \qquad A_y=(1/2i)(A_+-A_-)\ , \qquad
A_z=A_0 \ .
\end{equation}
In order to rewrite ${\hat {\mib M}}$ in another form, which 
will be convenient for later discussion, we introduce the following operators :
\begin{subequations}\label{7-9}
\begin{eqnarray}
& &{\hat {\mib r}}=({\hat x},\ {\hat y},\ {\hat z}) \ , \qquad 
({\hat r}=\sqrt{{\hat x}^2+{\hat y}^2+{\hat z}^2}) \ , 
\label{7-9a}\\
& &{\hat x}=2(x_+x_- - y_+y_-)\ , \quad
{\hat y}=2(x_+y_- + y_+x_-)\ , \quad
{\hat z}=x_+^2+y_+^2-x_-^2-y_-^2 \ , \nonumber\\
& &\qquad\qquad\qquad\qquad\quad ({\hat r}=x_+^2+y_+^2+x_-^2+y_-^2) \ , 
\label{7-9b}
\end{eqnarray}
\end{subequations}
\vspace{-0.6cm}
\begin{subequations}\label{7-10}
\begin{eqnarray}
& &{\hat {\mib p}}=({\hat p}_x,\ {\hat p}_y,\ {\hat p}_z) \ , 
\label{7-10a}\\
& &{\hat r}{\hat p}_x=(1/2i)(x_-\partial_{x_+}-y_-\partial_{y_+}
+x_+\partial_{x_-}-y_+\partial_{y_-}) \ , 
\nonumber\\
& &{\hat r}{\hat p}_y=(1/2i)(y_-\partial_{x_+}+x_-\partial_{y_+}
+y_+\partial_{x_-}+x_+\partial_{y_-}) \ , 
\nonumber\\
& &{\hat r}{\hat p}_z=(1/2i)(x_+\partial_{x_+}+y_+\partial_{y_+}
-x_-\partial_{x_-}-y_-\partial_{y_-}) \ . 
\qquad\qquad\qquad\quad\qquad\qquad
\label{7-10b}
\end{eqnarray}
\end{subequations}
It is easily verified that the operators ${\hat {\mib r}}$ and 
${\hat {\mib p}}$ obey 
\begin{eqnarray}\label{7-11}
& &[\ {\hat x}\ , \ {\hat p}_x\ ]=[\ {\hat y}\ , \ {\hat p}_y\ ]=
[\ {\hat z}\ , \ {\hat p}_z\ ]=i \ , \nonumber\\
& &[\ {\rm any\ other\ combination}\ ]=0 \ .
\end{eqnarray}
After rather tedious calculation, we obtain the form 
\begin{equation}\label{7-12}
{\hat {\mib M}}=-\omega^{-1}
\left[({\hat {\mib p}}\times {\hat {\mib L}})-
({\hat {\mib L}}\times {\hat {\mib p}})\right]+\left({\hat {\mib r}}/{\hat r}
\right)\cdot (2\omega)^{-1}{\hat {\cal H}}_4 \ . 
\end{equation}
Afterward, we can show that the form (\ref{7-12}) is closely related to 
the Runge-Lenz-Pauli vector in hydrogen atom. 

The eigenvalue equation (\ref{7-1}) is easily solved. For example, 
$\ket{\Psi}=\ket{\lambda\kappa;ll_0}$ shown in the form (\ref{3-3b}) is 
its eigenstate and the eigenvalue is given as 
\begin{equation}\label{7-13}
{\cal E}=2\omega(\lambda+\kappa+1) \ .
\end{equation}
Of course, $(\lambda, \kappa, l, l_0)$ obeys the rules shown in the relations 
(\ref{3-2}) and (\ref{3-3a}). 
The vacuum $\ket{0}$ satisfies the condition 
\begin{equation}\label{7-14}
{\hat c}_+\ket{0}={\hat d}_+\ket{0}={\hat c}_-\ket{0}={\hat d}_-\ket{0}=0 \ .
\end{equation}
With the use of the variables $x_{\pm}$ and $y_{\pm}$, 
$\ket{0}$ is expressed in the form 
\begin{equation}\label{7-15}
\ket{0}=\exp\left(-(\omega/2)(x_+^2+y_+^2+x_-^2+y_-^2)\right) \ . 
\end{equation}
In order to obtain $\ket{\lambda\kappa;ll_0}$ in a familiar form, we introduce 
the variables $(r,\phi, \theta,\psi)$ as follows : 
\begin{subequations}\label{7-16}
\begin{eqnarray}
& &x_+=\sqrt{r}\cos(\theta/2)\cos((\phi+\psi)/2) \ , \quad
y_+=\sqrt{r}\cos(\theta/2)\sin((\phi+\psi)/2) \ , \nonumber\\
& &x_-=\sqrt{r}\sin(\theta/2)\cos((\phi-\psi)/2) \ , \quad
y_-=\sqrt{r}\sin(\theta/2)\sin((\phi-\psi)/2) \ . \qquad
\label{7-16a}
\end{eqnarray}
Inversely, we have 
\begin{eqnarray}\label{7-16b}
& &r=x_+^2+y_+^2+x_-^2+y_-^2\ , \nonumber\\
& &\theta=\cos^{-1}\left(
\frac{x_+^2+y_+^2-x_-^2-y_-^2}{x_+^2+y_+^2+x_-^2+y_-^2}\right) \ , \nonumber\\
& &\phi=\tan^{-1}(y_+/x_+)+\tan^{-1}(y_-/x_-) \ , \nonumber\\
& &\psi=\tan^{-1}(y_+/x_+)-\tan^{-1}(y_-/x_-) \ . 
\end{eqnarray}
\end{subequations}
Here, new variables obey $0\leq r$, $0\leq \phi < 2\pi$, $0\leq \theta <\pi$ 
and $0\leq \psi <2\pi$. 
The vacuum $\ket{0}$ and the volume element $dV_4$ are expressed as 
\begin{equation}\label{7-17}
\ket{0}=\exp(-\omega r/2)\ , \qquad 
dV_4=rdr\sin\theta d\theta d\phi d\psi \ .
\end{equation}
The part $({\hat d}_-^*)^{2l}\ket{0}$ in the expression (\ref{3-3b}) is 
written in the form 
\begin{eqnarray}\label{7-18}
({\hat d}_-^*)^{2l}\ket{0}&=&
\left(\sqrt{\omega}(x_+-iy_+)\right)^{2l}\ket{0} \nonumber\\
&=&(\omega r)^{l}\exp(-\omega r/2)(\cos(\theta/2))^{2l}
\exp(-il(\phi+\psi)) \nonumber\\
&=&(\omega r)^l\exp(-\omega r/2)D_{-l,-l}^{(l)}(\phi\theta\psi) \ .
\end{eqnarray}
Here, $D_{-l,-l}^{(l)}(\phi\theta\psi)$ denotes the $D$-function appearing 
in the $so(3)$-group. 
We adopt the definition of the $D$-function in the textbook by 
Edmonds.\cite{9} 
We can see that $(\phi,\theta,\psi)$ denotes the Euler angle. 
In order to express other parts in the state (\ref{3-3b}), we must 
rewrite ${\hat L}_{\pm,0}$, ${\hat R}_{\pm,0}$ and ${\hat T}_{\pm,0}$ 
in terms of $(r,\phi,\theta,\psi)$. 
The results are as follows : 
\begin{eqnarray}
& &{\hat L}_{\pm}=e^{\pm i\phi}\left[
\pm\partial_{\theta}+i\cot\theta\ \partial_{\phi}-
i(\sin\theta)^{-1}\partial_{\psi}\right] \ , \nonumber\\
& &{\hat L}_0=-i\partial_{\phi} \ , 
\label{7-19}\\
& &{\hat R}_{\pm}=e^{\mp i\psi}\left[
\pm\partial_{\theta}-i\cot\theta\ \partial_{\psi}+
i(\sin\theta)^{-1}\partial_{\phi}\right] \ , \nonumber\\
& &{\hat R}_0=-i\partial_{\psi} \ , 
\label{7-20}\\
& &{\hat T}_{\pm}=-(1/4)\omega r-\omega^{-1}
\left[r\partial_r^2+(1\pm 1)\partial_r -{\hat {\mib L}}^2/r\right] 
\pm (1+ r\partial_r) \ , \nonumber\\
& &{\hat T}_0=(1/4)\omega r-\omega^{-1}\left[
r\partial_r^2+2\partial_r-{\hat {\mib L}}^2/r\right] \ .
\label{7-21}
\end{eqnarray}
We can see that ${\hat L}_{\pm,0}$ and ${\hat R}_{\pm,0}$ denote the 
angular momentum operators in the space- and the body-fixed frame, 
respectively, for rigid body. 
Therefore, we have 
\begin{eqnarray}\label{7-22}
({\hat L}_+)^{l+l_0}({\hat R}_-)^{\lambda-\kappa+l}({\hat d}_-^*)^{2l}\ket{0} 
=(\omega r)^l\exp(-\omega r/2)D_{\lambda-\kappa, l_0}^{(l)}(\phi\theta\psi) \ .
\end{eqnarray}
Further, for the Laguerre polynomial, we note the relation 
\begin{eqnarray}\label{7-23}
{\hat T}_+L_n^{2l+1}(\rho)\rho^l\exp(-\rho/2)D_{mm'}^{(l)}(\phi\theta\psi)
=(n-l)L_{n+1}^{2l+1}(\rho)\rho^l\exp(-\rho/2)D_{mm'}^{(l)}(\phi\theta\psi) \ .
\nonumber\\
& &
\end{eqnarray}
With the successive use of the relation (\ref{7-23}) from lower $n$, we obtain 
\begin{eqnarray}\label{7-24}
\ket{\lambda\kappa;ll_0}=
L_{\lambda+\kappa+1}^{2l+1}(\omega r)(\omega r)^l\exp(-\omega r/2)
D_{\lambda-\kappa,l_0}^{(l)}(\phi\theta\psi) \ .
\end{eqnarray}
The above is a possible expression of the eigenstate of the 
Schr\"odinger equation (\ref{7-1}) with the eigenvalue (\ref{7-13}). 
In the present case, it must be noted that we should use the 
following condition for the orthogonality : 
\begin{eqnarray}\label{7-25}
\int_0^{\infty}\rho^l\exp(-\rho/2)L_{n}^{2l+1}(\rho)\cdot \rho^l\exp(-\rho/2)
L_{n'}^{2l+1}(\rho)\cdot \rho d\rho=0\ . \quad (n\neq n')\qquad
\end{eqnarray}

Our final problem is to investigate the hydrogen atom in the framework of the 
present formalism. 
The hydrogen atom is described in three dimensional space and we use 
$(x,y,z)$ as the variables. 
The Schr\"odinger equation is written down as 
\begin{eqnarray}
& &{\hat {\cal H}}_3^0\kket{\Phi}=E\kket{\Phi} \ , 
\label{7-26}\\
& &{\hat {\cal H}}_3^0=-(1/2)(\partial_x^2+\partial_y^2+\partial_z^2)
-G/\sqrt{x^2+y^2+z^2} \ . 
\label{7-27}
\end{eqnarray}
Here, $G$ represents a positive constant, which characterizes the 
Coulomb potential. In order to analyze the above system in the four 
dimensional space, we define the variables $(x,y,z)$ by regarding them 
as $({\hat x},{\hat y},{\hat z})$ defined in the relation (\ref{7-9b}) : 
\begin{subequations}\label{7-28}
\begin{eqnarray}\label{7-28a}
& &{\hat x}=x=2(x_+x_- - y_+y_-) \ , \nonumber\\
& &{\hat y}=y=2(x_+y_- + y_+x_-) \ , \nonumber\\
& &{\hat z}=z=x_+^2+y_+^2-x_-^2-y_-^2 \ . \nonumber\\
& &\qquad (\sqrt{x^2+y^2+z^2}=x_+^2+y_+^2+x_-^2+y_-^2=r)
\end{eqnarray}
Associating the above, $\psi$ is adopted as the fourth, which is given in the 
relation (\ref{7-16b}) : 
\begin{equation}\label{7-28b}
\psi=\tan^{-1}(y_+/x_+)-\tan^{-1}(y_-/x_-) \ .
\end{equation}
\end{subequations}
The relations (\ref{7-28}) and (\ref{7-16a}) give us 
\begin{subequations}\label{7-29}
\begin{eqnarray}\label{7-29a}
& &{\hat x}=x=r\sin\theta\cos\phi \ , \nonumber\\
& &{\hat y}=y=r\sin\theta\sin\phi \ , \nonumber\\
& &{\hat z}=z=r\cos\theta \ , \nonumber\\
& &\psi=\psi \ . 
\end{eqnarray}
Further, the relation (\ref{7-10b}) leads us to 
\begin{eqnarray}\label{7-29b}
& &{\hat p}_x=-i\partial_x-iyz/r(x^2+y^2)\cdot \partial_\psi \ , \nonumber\\
& &{\hat p}_y=-i\partial_y+ixz/r(x^2+y^2)\cdot \partial_\psi \ , \nonumber\\
& &{\hat p}_z=-i\partial_z \ . 
\end{eqnarray}
\end{subequations}
We also obtain the relation 
\begin{eqnarray}\label{7-30}
& &(\partial_x^2+\partial_y^2+\partial_z^2)+(x^2+y^2)^{-1}
[\partial_\psi-2(x\partial_y-y\partial_x)]\partial_\psi \nonumber\\
& &=(1/4r)\cdot(\partial_{x_+}^2+\partial_{y_+}^2
+\partial_{x_-}^2+\partial_{y_-}^2) \ .
\end{eqnarray}
Then, we define the following equation : 
\begin{eqnarray}
{\hat {\cal H}}_4^0\ket{\Phi}&=&E\ket{\Phi} \ , 
\label{7-31}\\
{\hat {\cal H}}_4^0&=&
-(1/2)\left[4(x_+^2+y_+^2+x_-^2+y_-^2)\right]^{-1}
(\partial_{x_+}^2+\partial_{y_+}^2
+\partial_{x_-}^2+\partial_{y_-}^2)\nonumber\\
& &-G/(x_+^2+y_+^2+x_-^2+y_-^2) \ . 
\label{7-32}
\end{eqnarray}
Equation (\ref{7-31}) is easily solved as a result of Eq.(\ref{7-1}). 
It is rewritten as follows : 
\begin{eqnarray}
{\wtilde {\cal H}}_4\ket{\Phi^0}&=&4G\ket{\Phi^0} \ , 
\label{7-33}\\
{\wtilde {\cal H}}_4&=&
-(1/2)(\partial_{x_+}^2+\partial_{y_+}^2
+\partial_{x_-}^2+\partial_{y_-}^2)
+(1/2)(-8E)(x_+^2+y_+^2+x_-^2+y_-^2) \ . \nonumber\\
& &
\label{7-34}
\end{eqnarray}
Equation (\ref{7-33}) can be regarded as a special case of the relation 
(\ref{7-1}) with the form (\ref{7-5}). 
The correspondence of $4G$ and $-8E$ with ${\cal E}=2\omega(\lambda+\kappa
+1)$ shown in the relation (\ref{7-13}) and $\omega^2$ appearing in the 
Hamiltonian (\ref{7-5}) give us 
\begin{equation}\label{7-35}
2\omega(\lambda+\kappa+1)=4G \ , \qquad 
-8E=\omega^2 \ . 
\end{equation}
The relation (\ref{7-35}) leads us to 
\begin{eqnarray}\label{7-36}
& &E=-(1/2)G^2/(\lambda+\kappa+1)^2 \ , \nonumber\\
& &\omega=2G/(\lambda+\kappa+1)\ (=\omega_{\lambda+\kappa+1}) \ .
\end{eqnarray}
The eigenstate $\ket{\lambda\kappa;ll_0}$ is given as follows : 
\begin{eqnarray}\label{7-37}
\ket{\lambda\kappa;ll_0}=L_{\lambda+\kappa+1}^{2l+1}
(\omega_{\lambda+\kappa+1}r)(\omega_{\lambda+\kappa+1}r)^l
\exp(-\omega_{\lambda+\kappa+1}r/2)
D_{\lambda-\kappa,l_0}^{(l)}(\phi\theta\psi) \ .\qquad
\end{eqnarray}
The volume element $dV'_4$ for the state (\ref{7-37}) is obtained 
by calculating Jacobian for the relation (\ref{7-29a}) : 
\begin{equation}\label{7-38}
dV'_4=dxdydzd\psi=r^2dr\sin\theta d\theta d\phi d\psi \ . 
\end{equation}
The form (\ref{7-38}) should be compared with the relation (\ref{7-17}). 
The orthogonality of the Laguerre polynomial is, in the present case, 
given by 
\begin{eqnarray}\label{7-39}
& &\int_0^{\infty}\rho^l\exp(-\rho/2n)L_{n}^{2l+1}(\rho/n)
\cdot \rho^l\exp(-\rho/2n')
L_{n'}^{2l+1}(\rho/n')\cdot \rho d\rho=0\ . \quad (n\neq n')\nonumber\\
& &
\end{eqnarray}

Under the above preparation, let us investigate the hydrogen atom. 
First, we pay attention to the subspace spanned by the set (\ref{7-37}) 
with the condition 
\begin{subequations}\label{7-40}
\begin{eqnarray}
& &\lambda=\kappa=(n-1)/2\ , \qquad (n=1,\ 2,\ 3,\cdots) 
\label{7-40a}\\
i.e.,\ \ & &\lambda+\kappa+1=n \ . 
\label{7-40b}
\end{eqnarray}
\end{subequations}
The quantities $\lambda$ and $\kappa$ are integers or half-integers, 
and then, we have the relation (\ref{7-40b}). 
Further, the condition (\ref{3-3a}) gives us 
\begin{equation}\label{7-41}
l=0,\ 1,\ 2,\cdots,\ n-1 \ .
\end{equation}
Therefore, $\ket{\lambda\kappa;ll_0}$ $(=\ket{n;ll_0})$ becomes 
\begin{equation}\label{7-42}
\ket{n;ll_0}=L_n^{2l+1}(\rho/n)\rho^l e^{-\rho/2n}Y_{ll_0}(\theta\phi) \ . 
\quad (\rho=2Gr)
\end{equation}
The above is identical with the well-known wave function of the 
hydrogen atom. 
The reason why we arrived at this conclusion is simple. 
Any state specified by the condition 
(\ref{7-40}) does not depend on the variable $\psi$. 
In this case, ${\hat {\cal H}}_4^0$ defined in the relation (\ref{7-32}) 
plays the same role as that of ${\hat {\cal H}}_3^0$ given in the relation 
(\ref{7-26}) through the form (\ref{7-30}) and (\ref{7-28a}) or 
(\ref{7-29a}). 
In the space spanned by the state (\ref{7-42}), 
$\omega=2\sqrt{-2E}$ coming from the relation (\ref{7-35}) and 
${\hat {\cal H}}_4={\wtilde {\cal H}}_4=4G$ coming from the 
relation (\ref{7-33}) are permissible to use. 
Then, the form (\ref{7-12}) is reduced to 
\begin{eqnarray}\label{7-43}
{\hat {\mib M}}=-\frac{1}{\sqrt{-2E}}\left[
(1/2)(({\hat {\mib p}}\times {\hat {\mib L}})-
({\hat {\mib L}}\times {\hat {\mib p}}))-G\cdot{\hat {\mib r}}/r
\right] \ . 
\end{eqnarray}
The relation (\ref{7-43}) is identical with the Runge-Lenz-Pauli vector.

In the case of the $so(3,1)$-algebra, the vector ${\wtilde {\mib M}}$ 
defined in the relation (\ref{5-1b}) can be expressed in the form 
\begin{equation}\label{7-44}
{\wtilde {\mib M}}=-\omega^{-1}
\left[({\hat {\mib p}}\times{\hat {\mib L}})
-({\hat {\mib L}}\times{\hat {\mib p}})\right]
+({\hat {\mib r}}/{\hat r})\cdot(2\omega)^{-1}{\wtilde {\cal K}}_4 \ , 
\end{equation}
\vspace{-0.8cm}
\begin{subequations}\label{7-45}
\begin{eqnarray}
{\wtilde {\cal K}}_4&=&
2\omega\cdot(1/2)\left[({\hat c}_+^*{\hat d}_-^* - {\hat c}_-^*{\hat d}_+^*)
+({\hat d}_-{\hat c}_+ - {\hat d}_+{\hat c}_-)\right] \qquad\qquad
\nonumber\\
&=&2\omega\cdot(1/2)({\hat T}_+ + {\hat T}_-)=2\omega{\hat T}^0 \ .
\label{7-45a}
\end{eqnarray}
The Hamiltonian ${\wtilde {\cal K}}_4$ can be rewritten as 
\begin{eqnarray}
{\wtilde {\cal K}}_4&=&
-(1/2)(\partial_{x_+}^2 + \partial_{y_+}^2 + \partial_{x_-}^2 
+\partial_{y_-}^2) 
-(1/2)\omega^2(x_+^2 + y_+^2 + x_-^2 + y_-^2) \ . \qquad\ \ \ 
\label{7-45b}
\end{eqnarray}
\end{subequations}
Here, we used the relation (\ref{7-3}). It should be noted that the sign of 
the second term on the right-hand side of the form (\ref{7-45b}) 
is opposite of that of ${\wtilde {\cal H}}_4$ appearing in the 
relation (\ref{7-5}). 
In the same sense as that in the relation (\ref{7-1}), we set up the 
eigenvalue equation 
\begin{equation}\label{7-46}
{\wtilde {\cal K}}_4 \rket{\Psi}={\cal F}\rket{\Psi} \ .
\end{equation}
We can treat Eq.(\ref{7-46}) in the form given in the relations 
(2$\cdot$10a) and (2$\cdot$17) in (B). As is clear from the 
form (\ref{7-45a}), ${\wtilde {\cal K}}_4$ is composed of the sum of 
two $su(1,1)$-spins. 
Therefore, we can also adopt the method presented in (B). 
Through this procedure, ${\cal F}$ and $\rket{\Psi}$ are obtained. 
Of course, ${\cal F}$ is continuous and $\rket{\Psi}$ is 
non-normalizable.

The present scattering problem also obeys the Schr\"odinger equation 
(\ref{7-31}) with the Hamiltonian (\ref{7-32}). 
In the present problem, $E$ is positive. 
Then, after rewriting Eq.(\ref{7-31}) to the form (\ref{7-33}) with the 
Hamiltonian (\ref{7-34}), we have 
\begin{equation}\label{7-47}
8E=\omega^2 \ . 
\end{equation}
Since the eigenvalue of ${\wtilde {\cal H}}_4$ is $4G$, together with 
the relation (\ref{7-47}), ${\wtilde {\mib M}}$ can be expressed as follows : 
\begin{equation}\label{7-48}
{\wtilde {\mib M}}=-\frac{1}{\sqrt{2E}}
\left[(1/2)(({\hat {\mib p}}\times {\hat {\mib L}})
-({\hat {\mib L}}\times {\hat {\mib p}}))-G\cdot{\hat {\mib r}}/{\hat r}
\right] \ . 
\end{equation}
The vector ${\wtilde {\mib M}}$ is nothing but the 
Runge-Lenz-Pauli vector for the scattering problem.

In this paper, we proposed a possible boson realization of the 
$so(4)$- and the $so(3,1)$-algebra in the Schwinger type, 
the $su(2)$-algebra of which is familiar to us. 
Further, we applied it to the problem of electron moving in the Coulomb 
field induced by proton. The bound state is nothing but the hydrogen atom 
and it is closely related to the $so(3,1)$-algebra. 
The Runge-Lenz-Pauli vector was derived in the framework of the present 
boson realizations of both algebras. 
As a future problem, it may be interesting to apply the present boson 
realization to the description of nuclear dynamics, for example, 
the dynamics induced by the simplified pairing plus quadrapole 
interaction or the damping phenomenon of oscillator in 
thermal circumstance.

\section*{Acknowledgements}

On the occasion of publishing this paper, the authors S.N., Y.T. and 
M.Y. should acknowledge to Professor J. da Provid\^encia, 
co-author of this paper, for his continuous guide to the studies of 
many-body physics, and further, for his repeated invitations to visiting 
Coimbra. 
In fact, the main part of this paper, including Ref.\citen{8}, 
was completed when the above three stayed in Coimbra at the same period 
in September of 2004.

\end{document}